\documentstyle[]{mn}

\input{psfig.sty}

\ifnfsstwo

\fi
 
\ifnfssone
  \newmathalphabet{\mathit}
    \addtoversion{normal}{\mathit}{cmr}{m}{it}
    \addtoversion{bold}{\mathit}{cmr}{bx}{it}

\fi
 
\ifoldfss

\fi
 
\loadboldmathitalic
\loadboldgreek
 
\def\cc{cm$^{-2}$}
\def\ccc{cm$^{-3}$}
\def\kms{\,km~s$^{-1}$}      % note leading thinspace
\def\Zsun{\thinspace\hbox{$\hbox{Z}_{\odot}$}}
\def\msun{\thinspace\hbox{$\hbox{M}_{\odot}$}}

\def\gal{galaxy}

\def\el{elliptical}

\def\ev{evolution}

\def\for{formation}

\title[Lyman $\alpha$ Systems and Hot Halos]
      {Lyman $\alpha$ Systems within Hot Galactic Halos}
\author[S. M. Viegas, A. C. S. Fria\c ca and R. Gruenwald]
       {Sueli M. Viegas, Am\^ancio C. S. Fria\c ca and
        Ruth Gruenwald\\
        $^1$Instituto Astron\^omico e Geof\'\i sico, USP,
        Av. Miguel Stefano 4200, 04301-904 S\~ao Paulo, SP, Brazil}

\pubyear{1998}
\begin{document} 

\maketitle

\begin {abstract}
A hot gas halo is predicted by chemodynamical models during  the early
evolution of spheroidal galaxies. 
Cold condensations, arising from thermal instabilities in the hot gas, 
are expected to be embedded in the hot halo.
In  the early phases of the galaxy ($t \la 1$ Gyr),
a strong X-ray and EUV emission is produced by the extended hot gas 
distribution,
ionizing the cold clouds.
This self-irradiating two-phase halo model successfully explains
several line ratios observed in QSO absorption-line systems,
and reproduces  the temperature distribution of Lyman-$\alpha$ clouds.

\end{abstract}


\section{Introduction}

A hot halo is predicted by chemodynamical models during  the early
 evolution of spheroidal galaxies (Fria\c ca \& Terlevich 1998). 
In the chemodynamical model a hydrodynamical code is coupled to 
a chemical evolution solver to follow the \ev\ of a \gal\ since the stage
of a gaseous proto\gal.
After a settling phase, a partial wind 
develops, with a cooling flow in the inner regions of the \gal,
and an outflow in the outer parts, giving rise to an
extended hot gas distribution with strong soft X-ray and EUV emission in 
the early phases of the galaxy ($t \la 1$ Gyr). 
In the {\it self-irradiating two-phase halo model}
for quasar absorption line systems, the absorbers are identified to
colder condensations, embedded in the hot gas halo,
which have been formed by  thermal instabilities,
and which are photoionized by the surrounding soft X-ray and EUV radiation 
field.
This model successfully explained Lyman limit systems (LLSs) 
in the line of sight of QSO 1700+6416,
which show strong C and O high ionization absorption lines 
as well as a strong He I line (Viegas \& Fria\c ca 1995).
More recently, the analysis of a number of LLSs has shown that this 
model can also explain these systems
and predicts a subsolar C/O abundance ratio, 
as found in old Galactic stars and in agreement
with the results from chemical evolution models 
(Viegas, Fria\c ca \& Gruenwald 1997).
Recent observations  have revealed the presence of  heavy elements 
in  Ly$\alpha$ systems (Songaila \& Cowie 1996). 
Photoionization models
assuming the metagalactic radiation field, a very low density
for the absorbers ($\simeq$ 10$^{-2}$ \ccc), and low chemical abundances
($\simeq$ 10$^{-2}$ \Zsun) can reproduce some of the observed 
features.  
However, a few problems remain (Gruenwald \& Viegas 1993;
Giallongo \& Petitjean 1994, hereafter GP) : (a) the ionization parameter
given by the metagalactic radiation reaching a region with density
of the order of 10$^{-2}$ \ccc\ would be too low ($U \simeq 3\times 10^{-3}$)
to explain all the C IV systems, (b) if a lower density is assumed,
the size of the absorber would be too large to be located within a halo, 
and (c) the temperature of the Ly$\alpha$ systems would be too high. 

In this paper, we analyse the Ly$\alpha$ systems within the scenario
of the self-irradiating two-phase halo model, i.e., we assume that the absorbers 
are
the cold and dense clouds, originating from thermal instabilities
arising in the hot gas halo,
enriched by the strong wind in the early phases of the galaxy
evolution, and  photoionized by the hot gas radiation, 
which is the dominant ionizing radiation field.
In Section 2 the main characteristics of the 
absorbers derived from the models are discussed. 
In Section 3,  it is shown that
temperatures deduced from the absorption line profiles are reproduced
by our models. 
The conclusions are outlined in Section 4.

\section{Photoionization of the absorbers}

Before discussing the hot halo models, we give a brief account of 
the results  obtained by photoionization models
in which the ionization source is the metagalactic
radiation. 

\subsection{ Metagalactic ionizing radiation}

Photoionization models have been used for many years to analyse the 
QSO aborption systems assuming the metagalactic radiation
as the ionization source (Gruenwald \& Aldrovandi 1985, 
Bergeron \& Stasinska 1986). More recently, models considering a 
power law spectrum for the metagalactic flux and ionization parameter
in the range $10^{-2.4} \leq U \leq 10^{-1.5}$ have been invoked
to explain  the C II/C IV column density ratio deduced from
observed Ly$\alpha$ systems (Songaila \& Cowie 1996). 
The ionization balance of highly ionized species is sensitive to
the shape of the radiation spectrum, which is uncertain because of the
He$^+$ absorption in the IGM.  
Songaila and Cowie have then compared the
 observed ratios to models with an ionizing radiation spectrum
exhibiting a break as invoked
by GP to explain the temperature of Ly$\alpha$
 systems. However, in order to reproduce the  observed equivalent widths, a
high value for the ionization
parameter,  $U> 10^{-2}$, is required (Songaila \& Cowie 1996). 
It is generally assumed that  absorption systems have
densities $n_H \sim 10^{-2}$ cm$^{-3}$. 
For a metagalactic flux at the Lyman limit of the order
of 10$^{-21}$ ergs s$^{-1}$ cm$^{-2}$ Hz$^{-1}$ sr$^{-1}$,
these densities imply
$U \sim 10^{-3}$, which is much smaller than the values quoted above.
So, unless $n_H \ll 10^{-2}$ cm$^{-3}$ (which would imply
very large absorbers), the intensity of metagalactic radiaton field 
is too low to 
explain the observed systems and another energy source has to be 
invoked.
In addition, 
when deducing chemical abundances for Ly$\alpha$  and Lyman Limit
systems the presence of non-observed ions must be accounted for 
(Viegas 1995). This is usually done by  ionization correction factors 
which depend 
on the shape of the ionizing radiation spectrum. Thus, conclusions
about chemical abundances deduced 
from observed absorption lines based on photoionization models
should rely on models for which  the input parameters concerning the
ionizing radiation and the gas density are consistently chosen
from the comparison between theoretical and observed column densities.

In the following we discuss a consistent self-irradiating two-phase halo model 
for the
Ly$\alpha$ systems.

\subsection{ The self-irradiating two-phase halo}

As shown by the chemodynamical model of Fria\c ca \& Terlevich  (1998), 
in the early phases of an \el\ galaxy
evolution ($t \la 1$ Gyr), the extremely high rate of star-formation, 
which is converting
the initially entirely  gaseous proto\gal\ into the \gal\ stellar body,
leads to the \for\  of  a hot gas halo extending out to $\sim 100$ kpc
(for a present-day $\sim L^*$ \gal). 
This extended hot
gas distribution provides a soft X-ray and EUV radiation field, with
a 0.5-4 keV luminosity reaching a maximum of about 10$^{44}$
erg s$^{-1}$ at an age of a few $10^8$ yr, for a galaxy with
a total mass of $2\times$ 10$^{11}$ \msun. 
The hot gas in the halo is subjected to thermal instabilities, 
giving rise to cold clouds embedded in the hot gas. 
In this self-irradiating two-phase halo model, the dominant species of 
C, N, and O in the hot gas are He-like and H-like ions, 
whereas in the cold clouds less ionized species are present.  
Since the privileged epoch for formation of  large spheroids
seems to be at redshift $2<z <10$,
corresponding to a cosmic age interval $0.36-2.5$ Gyr
(for a $q_0=0.5$, $H_0=50$ \kms\ Mpc$^{-1}$ cosmology),
the $\sim 1$ Gyr length of the hot halo phase
makes it quite probable to find at high redshift ($z \ga 2$)
a young  \el\ galaxy in the hot halo stage.
At this redshift range, therefore, there is a large
supply of cold clouds inside galactic hot gas halos, that would act as
QSO absorbers.

Notice that a two-phase halo model has also been proposed by Giroux,
Sutherland \& Shull (1994) in order to explain the simultaneous 
presence of HI, HeI, and OIII  and the highly ionized species OVI  on the
line of sight of the QSO HS1700+6416. In their model, the OVI line
originates in a hot gas in collisional ionization,
while the low ionization species are produced in cold clouds.
In this case,  the  observed column
density of OVI requires a temperature for the hot gas of about
$2\times 10^5$ K.
However, in Fria\c ca \& Terlevich (1998) chemodynamical  models 
for elliptical galaxies, with masses spanning from $5 \times 10^{10}$
to $5 \times 10^{12}$ M$_{\odot}$, the halo is considerably more 
ionized (O VII and O VIII are the dominant ionization stages) 
and hotter ($T>$ a few times $10^6$ K). 
The evolutionary models show that when the
hot gas in the halo begins to cool down, the temperature rapidly falls
to a few times 10$^4$ K, the phase assumed by Giroux et al. being 
highly unstable. 
In our model (Viegas \& Fria\c ca 1995) for the LLSs 
at $z=2.1678$ and $z=2.433$ towards the QSO HS1700+6416, 
the OVI absorption
line originates in the outer layers of the cold clouds ionized
by the hot halo radiation.

Another important result from the chemodynamical model of galactic
evolution is the chemical enrichment history
of the galaxy stellar population as well as of the
hot gas halo. The fiducial model indicates that at an evolution time of the 
order of 0.3 Gyr, the C/H and O/H abundance ratios in the halo, relative to the
solar value, are of the order of 0.1 and 0.2, respectively.
 These values are higher than generally assumed for the
absorbers and are important because C and O are the main coolants 
of the gas, and also  because both  C/H and the ionizing radiation will set
the observed C IV/H I column.

In this paper, numerical simulations of absorbers are obtained with the 
photoionization code {\sc aangaba},  which  has been properly checked
against similar codes (Ferland et al. 1995). 
The metagalactic QSO radiation field
(Madau 1991) is added to the hot halo radiation field.
However, for a galaxy age $t \la 1$ Gyr, the effect of 
the metagalactic radiation is usually negligible. 
Following the results of the chemodynamical model, 
we assume for the halo gas a chemical abundance  of  0.1 solar. 

The spectrum shape of the hot gas radiation field evolves with time 
and depends on  the distance from the galactic centre (Viegas 
\& Fria\c ca 1995). 
The impact parameters associated with the QSO
absorption-line systems are a few Holmberg radii (Steidel 1993).
Thus, in order to analyse the Ly$\alpha$ cloud observations
we assume that the clouds, giving rise the absorption lines, 
are  at $r = 30$ or 100 kpc, for  evolution times of  $t = 0.21$ and 
0.36 Gyr, characteristic of the hot halo phase  for
the $M_G=2\times 10^{11}$ M$_{\odot}$ fiducial model of
Fria\c ca \& Terlevich (1998). 
The evolution of the model determines the temperature and density of the
hot gas, giving gas pressures
at the cloud edge in the range $200 < n_H T < 4000$ cm$^{-3}$ K.
Assuming pressure equilibrium in the cloud,  
the densities inferred from the photoionization calculations are generally 
larger than 
the value 10$^{-2}$ \ccc\ usually assumed for absorbers ionized by the
integrated QSO radiation field. 

\begin{figure}
 \centerline{
 \psfig{figure=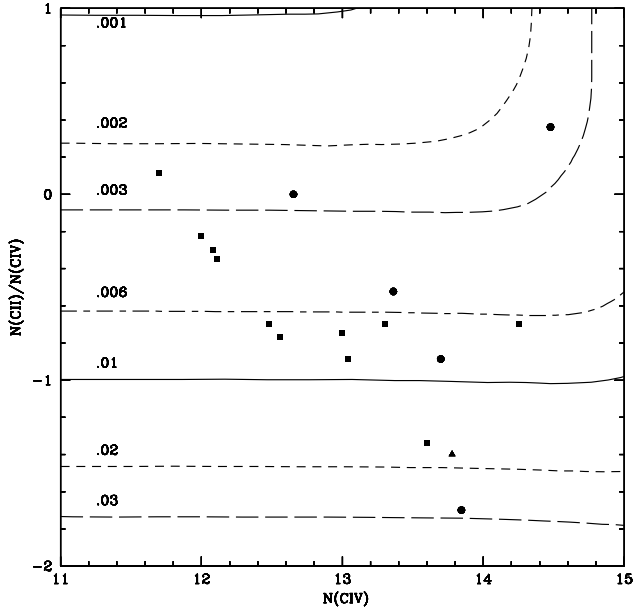,width=8.5cm,angle=0}}
 \caption{
N(C II)/N(C IV) as a function of N(C IV). The curves 
correspond to models and are labeled by the $U$ value. The
observational data come from Songaila \& Cowie (1996) and
correspond to systems with N(H I) $> 5\times 10^{14}$ cm$^{-2}$ 
at $3.135 < z < 3.60$ toward Q1422+23 (squares);
to systems with $N(H I) > 1.5\times 10^{15}$ cm$^{-2}$  at $z > 2.95$
toward Q0014+813 (triangles); 
and to partial Lyman limit systems 
($10^{17} < $N(H I)$ < 6\times 10^{17}$ \cc)
at $2.534<z<3.381$ toward several QSOs (dots).
}
\end{figure}

In order for our model to be consistent, the first step is to obtain 
the range of the ionization parameter. 
The value of $U$ can be inferred
from the ratio N(C II)/N(C IV), which depends only on 
the ionizing radiation spectrum. The theoretical results are 
compared to the observations (Songaila \& Cowie 1996) in 
the N(C II)/N(C IV) versus N(C IV) diagram (Figure 1).
For the observed sample, $0.002 \leq U \leq 0.03$, 
with 3$\times$ 10$^{11}$ $<$ N(C IV)
$<$ 4$\times$ 10$^{14}$ cm$^{-2}$.  

The C/H abundance 
can be obtained from the N(C~ IV)/N(H~I) versus N(H~I) diagram, since a change
in the abundance used in the model would lead to a vertical 
shift of the theoretical results.
Most of the observed results (Figure 2)
are explained by the models with C/H = 0.1 solar, including some Lyman 
limit systems (N(H I) $> 10 ^{17}$ cm$^{-2}$). 
However. some absorption systems in Figure 2 
require $U$ between 0.002 and 0.001, for C/H $=0.1$ solar.
Since a minimum value $U \approx 0.002$ is given by 
the N(C II)/N(C IV) ratio of the Lyman $\alpha$ systems (Figure 1),
the low N(C IV)/N(H I) objects may have 0.02 $<$ C/H $<$ 0.1 solar.

On the other hand, the Si/C abundance ratio can be obtained from 
the N(Si IV)/N(C IV) versus N(C II)/N(C IV) diagram (Figure 3).
The theoretical results were obtained with solar Si/C, 
and fit 50\% of the systems.
The remaining systems with a larger N(Si IV)/N(C IV), which could be
explained by a  supersolar Si/C abundance ratio, as indicated by the
long-dashed curve in Figure 3, corresponding to Si/C twice solar, for the
same  N(C II)/N(C IV) value. This fact has already been pointed out
by Songaila \& Cowie (1996). Notice, however, that in order
to reproduce all the N(Si IV)/N(C IV) ratios, in addition to
a Si/C overabundance, these authors also need to
assume an ionizing radiation spectrum largely absorbed at the He$^+$ 
threshold, as invoked by GP to explain
the temperature of the Lyman $\alpha$ clouds.

\begin{figure}
 \centerline{
 \psfig{figure=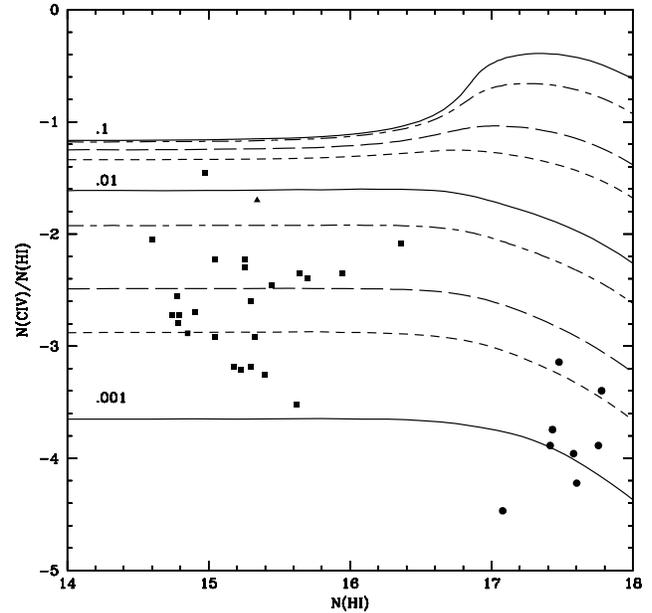,width=8.5cm,angle=0}}
 \caption{
N(C IV)/N(H I) versus N(H I). The notation is
the same as in the Figure 1.
}
\end{figure}

\section{The Temperature Puzzle}

As pointed out by GP, previous
observations of Lyman $\alpha$ systems 
with resolution of 20 to 30 km s$^{-1}$ indicated an average
value of the Doppler width of about 30 km s$^{-1}$. The inferred
temperature of the gas is then less than $5\times 10^4$ K. This temperature
could be explained by photoionization of a 
low density gas ($\la 10^{-3}$ cm$^{-3}$)
by the integrated QSO radiation field.
Higher resolution observations (Pettini et al. 1990, Carswell et al. 1991,
Hunstead \& Petini 1991, Rauch et al. 1992, 1993, Giallongo et al. 1993)
revealed, however, narrower absorption lines
indicating lower temperatures ($T \simeq 2\times 10^4$ K), which
can not be explained by the above models. 
In order to obtain
low temperatures in a low abundance and dilute gas, GP
proposed that the ionizing spectrum should be absorbed at the He$^+$ Lyman limit
(at 4 Ryd), thus decreasing the heating energy due to photoionization. 
However, GP need to vary the amplitude of the spectrum break at 4 Ryd 
by a factor of $10^3$ to obtain temperatures in the range $2 - 5 \times 10^4$ K, 
as required by the determinations of  both low and high temperatures 
for Lyman $\alpha$ systems.
In addition, in order to have an ionization 
parameter high enough to assure a highly ionized gas, the density 
of the clouds must  be lower than $10^{-3}$ cm$^{-3}$, leading to a
recombination time close to the Hubble time.

\begin{figure}
 \centerline{
 \psfig{figure=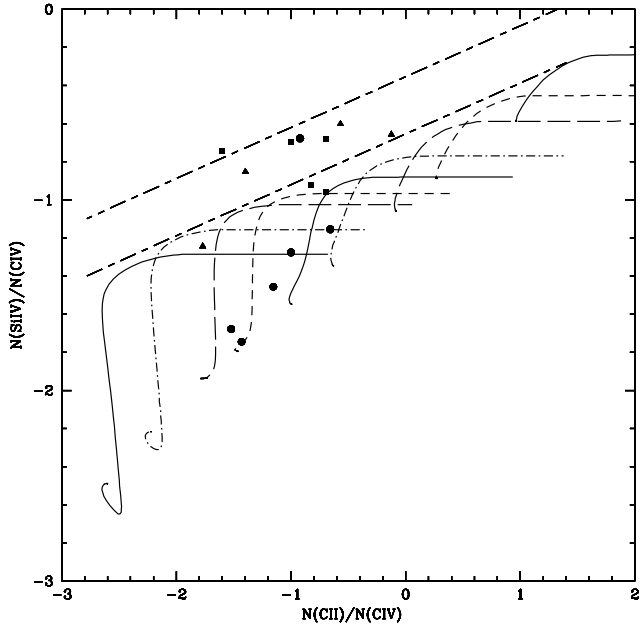,width=8.5cm,angle=0}}
 \caption{
N(Si IV)/N(C IV) versus N(C II)/N(C IV). The curves
correspond to the models shown in Figures 1 and 2. The lower
straight line (long-short dashed) defines the limit given
by the models with solar Si/C ratio, while the upper
straight line corresponds to twice solar.
}
\end{figure}

In the self-irradiating two-phase halo model, the absorbers are associated with
relatively cold and dense clouds embedded in the extended hot gas halo, 
and the radiation from the hot phase gas maintains the high
ionization state of the Ly$\alpha$ clouds as shown in 
Figures 1 and 2. 
For these clouds, the values  of the
ionization parameter indicated by N(C II)/N(C IV) ratio range from 
0.002 to 0.03, and the temperature from $10^4$
to $5\times 10^4$ K (Figure 4, bottom panel). 
In the figure,
the variation of  temperature with $U$ is shown for chemical abundances
0.01 and 0.1 solar. 
The temperature range obtained from the
observed absorption line profiles (Giallongo et al. 1993)
is indicated by the solid vertical 
bar, whereas the temperature range obtained 
from photoionization models with the integrated QSO
radiation field without a deep break at 4 Ryd (GP)
is indicated by the dashed vertical bar.
The results of the self-irradiating two-phase halo are in very good agreement
with the observed temperatures. 
In addition, the characteristic 
size of filaments with N(H I) = 10$^{14}$ - 10$^{15}$ 
cm$^{-2}$ is 0.1 to 100 pc (Figure 4, top panel), being
almost independent of the chemical abundance of the gas.  
For a given N(H I), 
the higher the temperature, the larger the filament.
Notice that the inverse correlation is found in GP 
for models with Compton cooling and a break of the ionizing spectrum
at 4 Ryd.
Observations of the C~II$^*$ absorption line 
could provide an estimate of the absorber volumetric density .
Therefore, the real size of the absorbers is unknown, and
the relation between the absorber size and its temperature
can not be tested yet. 
The difference between the two models is illustrated even more dramatically
by the absorber mass, which is in the range 
$10^{7} - 10^{10}$ \msun\ in GP, while in our model
it is $10^{-3}$ to 2  \msun, for N(H I) = 10$^{14}$ to 10$^{15}$ 
cm$^{-2}$, if a spherical symmetry is adopted.  However, if the halo
condensations are produced by shocks, it is likely that they would look like
sheets of gas with much larger masses. 

As pointed out previously, the hot gas is highly ionized and contributes
only to the column density of the He-like and H-like ions of  CNO.
Thus, the temperature of the Ly$\alpha$  systems discussed above
refers only to the clouds.

\begin{figure}
 \centerline{
 \psfig{figure=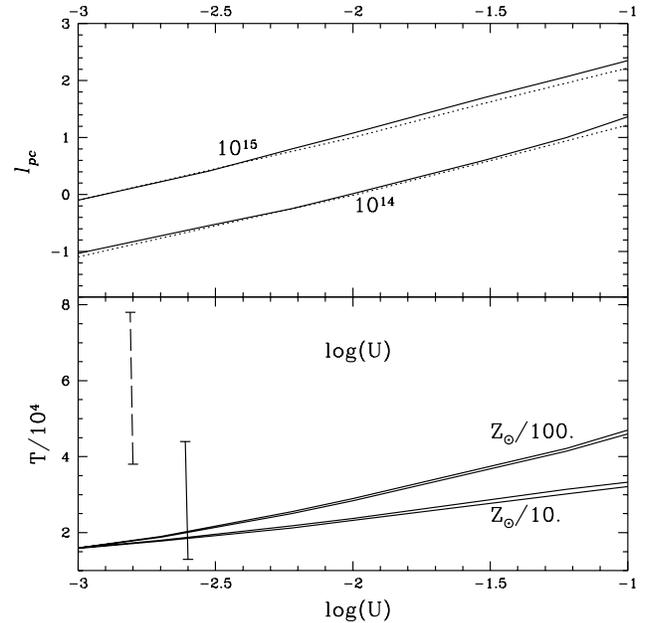,width=8.5cm,angle=0}}
 \caption{
Model results for the temperature (bottom panel)
and the size (top pannel) of the Ly $\alpha$ systems as a function
of the ionization parameter $U$. 
Top panel: the results are given 
for N(H I) $= 10^{14}$ cm$^{-2}$ and $10^{15}$ cm$^{-2}$, 
for (C/H)/(C/H)$_{\odot}$ $=0.1$ (solid line) and 0.01 (dotted line). 
Bottom pannel: 
the curves given by the models are labeled by the value of C/H.
In each case the results are shown for 
N(H I) $= 10^{14}$ cm$^{-2}$ and $10^{15}$ cm$^{-2}$.
The solid vertical bar indicates the range of the temperature
of observed systems while the dashed vertical bar corresponds
to the range covered by the models of Giallongo and Petitjean
(1994) with the integrated QSO radiation field.
}
\end{figure}

\section{Conclusions}

We have shown that a self-irradiating two-phase halo,
which is suggested by a chemodynamical model of galactic
evolution,  can explain the 
results concerning the presence of heavy
element absorption lines in Lyman $\alpha$ systems.
In particular, since this model allows densities
of the absorption systems higher ($\simeq 0.1$ \ccc)
than in the standard model ($\la 10^{-3}$ cm$^{-3}$),
the gas cooling is more efficient, leading to lower temperatures
in agreement with the observations. In addition, 
a higher density implies a smaller and less massive
absorber, where the cooling time is of the order of the
recombination time and much smaller than the Hubble time.
Therefore, photoionization models assuming ionization equilibrium 
are adequate to describe the Ly$\alpha$ absorbers.
Two results of our model for the Ly$\alpha$ absorbers may be
tested in the near future: the small size of the absorbers
and their low mass.

Another important point concerns the chemical abundance
of the absorbing gas. The chemodynamical model indicates
that the halo gas is enriched in heavy elements in a 
short time scale, reaching solar abundances 
by $t< 3$ Gyr, even at 100 kpc from the galactic center.
It is also in this 
evolutionary phase that the hot halo develops, so the 
absorbers may be more enriched than it is usually 
assumed. 
Our results indicate that most of the 
observed Lyman $\alpha$ absorbers, consistently analysed here using
a photoionization model, have C/H $\approx 0.1$ solar.
For a few others, the carbon abundance should be
a factor of 5 smaller. 
These results are in agreement
with the enrichment of the hot gas halo, predicted by the 
chemodynamical model. 
Regarding the Si abundances, our analysis of the Si IV absorption line also 
indicates
an abundance ratio  $1\la$ Si/C $\la 2$ solar (Figure 3).
Although the \ev\ of  the Si abundance
is not included in the present version of the chemodynamical  code,
its behaviour is similar to that of the other $\alpha$-elements, and, therefore,
can be traced by the \ev\ of the O.
Whereas the $\alpha$-elements are produced in massive stars
($M \ga 10$ \msun) via Type II+Ib supernova events,
the C is partially produced in Type II+Ib supernovae, and most of it
in intermediate-mass stars ($0.8 < M < 8$ \msun).
Its behaviour, therefore, is intermediate between that of  O (and Si)
and that of  Fe, mainly produced by Type Ia supernovae.
This explains the contrast between the typical overabundace of  O,
of order of [O/Fe]=+0.5, found by Gratton et al. (1998) for low-metallicity
inner halo and thick disk stars,
and the moderate C enrichment inferred for metal-deficient stars.
In fact, [C/Fe]=+0.2 at [Fe/H]=-0.8 is obtained by Tomkin et al. (1995) 
in their analysis of 105 dwarf stars, whereas Carbon et al (1987),
for a sample of 83 dwarfs in the range $-2.5 \leq$ [Fe/H] $\leq -0.6$,
found a solar [C/Fe] ratio over most of the metallicity range, with
a slight upturn of [C/Fe] at very low metallicities.
Therefore, we expect an increase of the $\alpha$-elements/C ratio from solar
to 2-3 solar, as the metallicity varies from solar to 0.1-0.01 solar.
This observational trend  is reproduced by the chemodynamical model,
which predicts O/C $\sim $ 2 solar for C/H $\sim$ 0.1 solar. 
Thus,  $1\leq$ Si/C $\leq2$ solar may be expected in the absorbers, 
in agreement with our results.

It should be noted that a fraction of the elements in the Lyman $\alpha$
absorbers is expected to be in the form of grains, thus reducing the gas
phase abundances, which are precisely those being measured
through the absorption lines.
Among the elements observed in QSO absorption-line systems,
O, S and Zn are almost unaffected by depletion into grains in the Galactic ISM,
whereas C, Si and Fe are typically reduced by 1-2 dex (Savage \& Sambach 1996).
Therefore, the observational points in Figure 3 could be subject to corrections
for depletion into grains.

\section*{Acknowledgments}
 
We acknowledge support from the Brazilian agencies CNPq,
FAPESP, and FINEP/PRONEX.

\bsp

\end{document}